\documentclass[epjCONF]{svjour}
\usepackage{graphics}
\usepackage[varg]{txfonts} 
\usepackage[latin1]{inputenc}
\session-title{Conference Title, to be filled}
\begin{document}
\title{Tracing early stellar evolution with asteroseismology: pre-main sequence stars in NGC 2264}
\author{Konstanze Zwintz\inst{1}\fnmsep\thanks{\email{konstanze.zwintz@ster.kuleuven.be}} \and Luca Fossati\inst{2} \and Tatiana Ryabchikova\inst{3} \and David Guenther\inst{4} \and Conny Aerts\inst{1,5} }
\institute{Institute of Astronomy, KU Leuven, Celestijnenlaan 200D, 3001 Leuven, Belgium \and Argelander Institute, University of Bonn, Germany, Auf dem H\"ugel 71, 53121 Bonn, Germany \and Institute of Astronomy, Russian Academy of Sciences, Pyatnitskaya Str 48, 119017 Moscow, Russia \and Department of Astronomy and Physics, St. Mary's University, Halifax,
    NS B3H 3C3, Canada \and Department of Astrophysics, IMAPP, Radboud University Nijmegen, P.O. Box 9010, 6500 GL Nijmegen, Netherlands}

\abstract{
Asteroseismology has been proven to be a successful tool to unravel details of the internal structure for different types of stars in various stages of their main sequence and post-main sequence evolution. Recently, we found a relation between the detected pulsation properties in a sample of 34 pre-main sequence (pre-MS) $\delta$ Scuti stars and the relative phase in their pre-MS evolution. With this we are able to demonstrate that asteroseismology is similarly powerful if applied to stars in the earliest stages of evolution before the onset of hydrogen core burning.
} 
\maketitle
\section{Introduction}
A wealth of different classes of pulsators is known among the main sequence (MS) and post-main sequence (post-MS) stars and studied in great detail using asteroseismic methods (see, e.g., Aerts et al. 2010 \cite{aer10}). 
The different groups depicted in the Hertzsprung-Russell (HR) diagram for stellar oscillations (Christensen-Dalsgaard 1988 \cite{chr88})  can be distinguished according to the different excitation mechanisms responsible for the oscillations (e.g., the $\kappa$-mechanism, convective blocking or stochastic excitation) and by the restoring forces, pressure or buoyancy, that cause either pressure (p; e.g., in $\delta$ Scuti stars) or gravity (g; e.g. in $\gamma$ Doradus stars) modes originating from different depths inside the stars (e.g., Aerts et al. 2010 \cite{aer10}). But the HR diagram for stellar oscillations includes stars of all evolutionary stages from the most evolved objects (e.g., pulsating white dwarfs), to post-MS stars (e.g., red giant oscillators or RR Lyrae stars) and MS pulsators (e.g., $\beta$ Cephei stars). Some of the regions close to the zero-age main sequence (ZAMS) contain either pulsating MS or pre-main sequence (pre-MS) stars, in particular for the $\delta$ Scuti, $\gamma$ Doradus and Slowly Pulsating B (SPB) star domains. Therefore, the HR diagram for stellar oscillations includes a mixture of evolutionary stages, and, hence, stellar ages.

The MS and post-MS pulsators have been studied for decades revolutionizing our understanding of galactic and stellar evolution. We can now show that asteroseismology of pre-MS stars has similar power for young stellar objects and will be able to contribute to a better understanding of the earliest phases of stellar evolution. 

\subsection{Pre-MS $\delta$ Scuti stars}
The first detailed investigation of oscillations in pre-MS stars only dates back in the late 1990ies where the first seismic study of the young $\delta$ Scuti type star HR 5999 was conducted (Kurtz \& Marang 1995 \cite{kur95}). At that time pre-MS $\delta$ Scuti stars were believed to be purely radial pulsators showing only few pulsation modes. Currently, the group of pre-MS $\delta$ Scuti pulsators is well established with $\sim$ 60 members (Zwintz et al. 2011 \cite{zwi11}) that have been proven to show radial and non-radial, acoustic (p-) mode pulsations with periods between ~15 minutes and 7 hours. 

\section{Determination of the evolutionary stage from asteroseismology}
Close to the ZAMS, the pre- and post-MS evolutionary tracks intersect (see, e.g., Breger \& Pamyatnykh 1998 \cite{bre98}) making the identification of the evolutionary stage of a given star from its fundamental parameters (i.e., effective temperature, $T_{\rm eff}$, gravity, log\,$g$, and mass) ambiguous. As pulsation modes carry information about stellar interiors and show a different pattern for stars in the pre- or the (post-) main sequence phase (Suran et al. 2001 \cite{sur01}), it is possible to use asteroseismology to constrain the evolutionary stage of a star (Guenther et al. 2007 \cite{gue07}). 

For the investigation of the pulsation properties of pre-MS stars and the connection to their relative evolutionary phase, several ingredients are needed.

\subsection{Stars in the pre-MS stage}

The pre-MS nature of a star cannot be identified unambiguously from its observational features. Indicators for stellar youth include the association of a star to a star forming region, the presence of infrared and / or ultraviolet excesses and emission lines or the classification as a Herbig Ae / Be star (Herbig 1962 \cite{her62}). But these features can also be misleading, as for example the comparatively old, low-mass asymptotic giant branch (AGB) and post-AGB stars show similar observational properties (e.g., infrared excess and emission lines) and populate the same region in the HR diagram as young stellar objects (Kamath et al. 2014 \cite{kam14}). This problem can be alleviated by the membership of a given star to an open cluster younger than $\sim$ 10 million years, where the most massive O and B stars have already started to burn hydrogen in their cores, but stars of later spectral types are still in their pre-MS evolutionary stage.

\subsection{Time-series to discover and analyze variability}

The first photometric time series for pre-MS $\delta$ Scuti type pulsators have been obtained through ground-based campaigns (e.g., Kurtz \& Marang 1995 \cite{kur95}; Zwintz et al. 2006 \cite{zwi06}). With the launch of the MOST (Walker et al. 2003 \cite{wal03}) and CoRoT (Baglin 2006 \cite{bag06}) satellites, photometric time series from space became available for the analysis of pre-MS pulsators (e.g., Zwintz et al. 2011 \cite{zwi11}). Using the ultra-precise data from space, we are challenged by the sometimes rather complex nature of pre-MS objects where irregular variability caused by circumstellar material and pulsational variability on the millimagnitude level can be present at the same time as in the case for HD 142666 (Zwintz et al. 2009 \cite{zwi09}).

\subsection{Position in the HR diagram}
The determination of the fundamental parameters -- $T_{\rm eff}$, log\,$g$, projected rotational velocity ($v$\,sin\,$i$), radial velocity and metallicity ([Fe/H]) -- of the pre-MS pulsators was conducted using mainly dedicated high-resolution, high signal-to-noise spectroscopy obtained with the 2.7m telescope at McDonald Observatory equipped with the Robert G. Tull \'echelle spectrograph, the ESO VLT with UVES, the ESO 3.6m telescope with HARPS, the Canada France Hawaiian Telescope (CFHT) with the spectropolarimeter ESPaDOnS and the 1.2m Mercator telescope with the HERMES \'echelle spectrograph. We complemented our analysis with few low resolution spectra using the R-C spectrograph at the Cerro Tololo Interamerican Observatory (CTIO) 1.5m telescope, data available in archives (e.g., the ESO archive) and some values reported in the literature.

\subsection{Pre-MS evolutionary tracks}
With the previously determined fundamental parameters, the positions of our selected sample of 34 pre-MS $\delta$ Scuti stars in the HR diagram can be investigated and compared to the theoretical pre-MS evolutionary tracks (Guenther et al. 2009, \cite{gue09}). It is evident that members of this sample are in different relative evolutionary phases before the ZAMS, hence they also have different ages. The 34 pre-MS $\delta$ Scuti stars show $v$\,sin\,$i$ values up to $\sim$190\,km$^{\rm -1}$ which corresponds to a maximum of 50\% of their breakup velocity (Zwintz et al. 2014 \cite{zwi14}).

\section{Asteroseismic Interpretation}
For the asteroseismic description of the observed pulsation properties, the highest observable p-mode frequency, $f_{max}$, was used. It is the observable counterpart of the theoretical acoustic cutoff frequency, i.e., the highest p-mode frequency that can be trapped inside a star (e.g., Aerts et al. 2010 \cite{aer10}). We could show that there is a clear relation between $f_{max}$ and the relative pre-MS evolutionary stage: the least evolved stars (i.e., closer to the birthline) are the slowest pulsators, while the most evolved stars (i.e., closest to the ZAMS or on the ZAMS) show the highest $f_{max}$ values (Zwintz et al. 2014 \cite{zwi14}). This clearly shows that it is possible to trace early stellar evolution with oscillations and illustrates the power of asteroseismology also for the pre-MS evolutionary stages. 

In particular the illustration of the stack of amplitude spectra of nine $\delta$ Scuti type pulsators in the young cluster NGC 2264 (see Zwintz et al. 2014 \cite{zwi14}, their Figure 3), resembles a lot the first results on the connection between the evolutionary stage and the pulsation properties for nine red giants (De Ridder et al. 2009 \cite{der09}, see their Figure 1). In both cases, the first space data came from the MOST and CoRoT satellites. For the red giant oscillators, the Kepler space telescope (Gilliland et al. 2010 \cite{gil10}) provided a giant leap in the asteroseismic interpretation, while no pre-MS stars were included in the main Kepler mission. 

\section{Future prospects}
The Space Photometry Revolution for young stellar objects has already started with data obtained by the MOST and CoRoT satellites. It will be continued with data expected from the K2 mission. In the future missions like TESS (Ricker et al. 2014 \cite{ric14}) and PLATO (Rauer et al. 2014 \cite{rau14}) will have the potential to bring the Space Photometry Revolution for pre-MS objects to a new level. 

As star formation and planet formation are directly linked to each other, understanding the physical processes that occur in these early stages of stars is essential. Although we have a general concept of how stars and planets are formed and evolve, our current knowledge of early stellar and planetary evolution is limited and contains a lot of unsolved questions. Studying oscillations in pre-MS objects has now been shown to be a powerful tool to answer some of these questions.

\begin{acknowledgement} 
The research leading to these results has received funding from the Research Council of the KU Leuven under grant agreement GOA/2013/012, and from the Fund for Scientific Research of Flanders (FWO), Belgium, under grant agreement G.0B69.13. LF acknowledges support from the Alexander von Humboldt foundation. TR acknowledges partial financial support from the Presidium RAS Program ``Nonstationary Phenomena in Objects of the Universe''. 
\end{acknowledgement}


\begin{thebibliography}{}
\bibitem{aer10} Aerts, C., Christensen-Dalsgaard, J., Kurtz, D. W., \textit{Asteroseismology} (Springer, Dordrecht Heidelberg London New York, 2010)
\bibitem{chr88} Christensen-Dalsgaard, J., \textit{Proceedings of the IAU Symposium No. 123 ``Advances in Helio- and Asteroseismology''}, edited by J. Christensen-Dalsgaard and S. Frandsen, (D. Reidel Publishing Co., Dordrecht, 1988), 295 
\bibitem{kur95} Kurtz, D. W., Marang, F., MNRAS, \textbf{276}, (1995) 191
\bibitem{zwi11} Zwintz, K., Kallinger, T., Guenther, D. B., et al, ApJ, \textbf{729}, (2011), 20
\bibitem{bre98} Breger, M., Pamyatnykh, A., A\&A \textbf{332}, (1998), 958
\bibitem{sur01} Suran, M., Goupil, M., Baglin, A., Lebreton, Y., Catala, C., A\&A \textbf{372}, (2001) 233
\bibitem{gue07} Guenther, D. B., Kallinger, T., Zwintz, K., Weiss, W. W., Tanner, J., ApJ \textbf{671}, (2007), 581
\bibitem{her62} Herbig, G. H., Advances in Astronomy and Astrophysics \textbf{1}, (1962), 47
\bibitem{kam14} Kamath, D., Wood, P. R., Van Winckel, H., MNRAS \textbf{439}, (2014) 2211
\bibitem{zwi06} Zwintz, K., Weiss, W. W., A\&A, \textbf{457}, (2006) 237
\bibitem{wal03} Walker,G.A.H., Matthews, J.M., Kuschnig, R., et al. PASP \textbf{115}, (2003) 1023
\bibitem{bag06} Baglin, A., \textit{The CoRoT Mission} (ESASP 1306, Noordwijk, The Netherlands: ESA Publications Division, 2006)
\bibitem{zwi09} Zwintz, K., Kallinger, T., Guenther, D. B., et al., A\&A, \textbf{494}, (2009) 1031
\bibitem{gue09} Guenther, D. B., Kallinger, T., Zwintz, K., ApJ \textbf{704}, (2007), 1710 
\bibitem{zwi14} Zwintz, K., Fossati, L., Ryabchikova, T., et al., Science \textbf{345}, (2014), 550
\bibitem{der09} De Ridder, J., Barban, C., Baudin, F., et al., Nature \textbf{459}, (2009), 398
\bibitem{gil10} Gilliland, R. L., Brown, T. M., Christensen-Dalsgaard, J., et al., PASP \textbf{122}, (2010) 131
\bibitem{ric14} Ricker, G. R., Winn, J. N., Vanderspek, R., et al., \textit{Proceedings of the SPIE}, \textbf{9143}, (2014) 20
\bibitem{rau14} Rauer, H., Catala, C., Aerts, C., et al., 2014, Experimental Astronomy, in press (arXiv:1310.0696)
\end{thebibliography}
\end{document}